# TRANSPARENT MIGRATION OF MULTI-THREADED APPLICATIONS ON A JAVA BASED GRID

T. N. Ellahi, B. Hudzia, L. McDermott and T. Kechadi
School of Computer Science and Informatics
University College Dublin
Dublin - Ireland
{benoit.hudzia, liam.mcdermott, tariq.ellahi, tahar.kechadi}@ucd.ie

**ABSTRACT**
Grid computing has enabled pooling a very large number of heterogeneous resource administered by different security domains. Applications are dynamically deployed on the resources available at the time. Dynamic nature of the resources and applications requirements makes needs the grid middleware to support the ability of migrating a running application to a different resource. Especially, Grid applications are typically long running and thus stoping them and starting them from scratch isn't a feasible option. This paper presents an overview of migration support in a java based grid middleware called DGET. Migration support in DGET includes multi-threaded migration and asynchronous migration as well.

## 1 Introduction

Grid systems[1][2] offer the capability of sharing resources at a very large scale across the organization boundaries. Applications can be deployed on a large number of resources to exploit more computational power and thus increasing the performance. Grid is inherently dynamic and volatile environment. Resources are added and removed frequently. Dynamic nature of the resources and applications requirements makes needs the grid middleware to support the ability of migrating a running application to a different resource. Especially, Grid applications are typically long running and thus stoping them and starting them from scratch isn't a feasible option.

In this paper, we present mechanism to support transparent strong migration of grid applications in a Java based grid. This mechanism is implemented in DGET grid middleware. The rest of the paper is structured as follows: related work is discussed in section 2. Section 2 presents the problem statement followed by an introduction of DGET in section 4. Details of migration mechanism is explained in section 5 with experimental results give in section 6. Section 7 gives the conclusion and future work.

## 2 Related Work

There are a number of projects that are developed in java and provide transparent strong migration facility. These projects can be divided in 4 categories. Following is a description of each category and an overview of the projects that followed the approach:

**Modified Java Virtual Machine**   This approach requires modifying the Java Virtual Machine. Using a modified JVM enables capturing and restoring execution state with zero overhead. JavaThread [3], D'Agents[4], Sumatra[5], NOMADS[6] and Ara[7] systems use a modified JVM to support strong migration capability. Using a modified JVM does not incur any performance overhead as exist in other approaches due to insertion of source code/bytecode to capture and restore the execution state. The drawback of using a modified JVM to support migration is lack of portability. The solution can not be deployed on any standard JVM.

**Source Code Transformation:**   Second approach pre-processes the application source code to add code blocks for capturing and restoring execution state. This approach results in huge space and time overhead thus degrading the application performance. Another drawback of using a source code preprocessor approach is that it is no practical to obtain source code of the application classes especially if the application relies on third party libraries. This approach is followed by WASP[8] and JavaGo[9] projects. In these systems, migration can only be self-invoked by the application at some specific point during its execution. It is not possible for an external entity to invoke migration operation on some application asynchronously to its execution. An example of this can be migration initiated by the load balancing process. In addition to the lack of support for asynchronous migration, multi-threaded migration isn't supported. OrganicGrid[10] project also relies on this approach. Despite supporting asynchronous migration and multi-threaded migration features, it suffers from the space and time overhead as each method of a class and transformed into a separate class.

**Bytecode Instrumentation:** Another approach for supporting migration capability is to post-process the application classes after compilation. This approach is similar to the previous one except that it works at lower level. Instead of modifying the application source code, application bytecode is transformed to add code blocks for capturing and restoring execution state. This approach is rather practical but suffers from the space and time overhead as well. Brakes[11] and JavaGoX[12] follow this approach and modify byte-code before executing the application. Both systems don't support asynchronous and multi-threaded migration.

**JPDA & Bytecode Instrumentation:** Last category is the approach where execution state is captured and restored using a combination of JPDA and bytecode instrumentation. Project CIA[13] and JESSICA[14] follow this approach. Our approach is most similar to CIA but CIA lacks asynchronous and multi-threaded migration. JESSICA supports thread migration within a cluster. Asynchronous migration is supported but it lacks support for multi-threaded migration.

## 3 Problem Statement: Transparent Strong Migration

In this section, we will give an overview of the reasons why migration support is required in grid systems. A set of requirements that must be supported by an effective and feasible migration solution will be laid out.

### 3.1 Why Migration Support is Required in Grid:

Migration support in grid systems is necessitated by a variety of reasons. Following are some of the most important reasons for grid middleware to support application migration:

**Data-driven Relocation:** Certain grid applications operate on distributed data resources for example distributed data mining applications. In order to exploit the locality and improve the performance, applications can be migrated closer to the data resources to improve performance.

**Resource Availability:** Grid systems exhibit dynamic behavior. Both grid resources and application requirements are dynamic and variable over time. When application requirements change or resources characteristics fall below the application expectations, new resources must be found and application should be migrated to these new resource that satisfy the application requirements.

**Dynamic Reconfiguration:** Resources in grid systems join and leave frequently. In case of anticipated resource departure, the computations initiated on the outgoing resources must be relocated to other resources in order to avoid any loss of work performed.

**Load Balancing:** Load balancing needs to be performed in order to move computations from highly loaded resources to the lightly loaded resources. Migration can help perform load balancing activities.

### 3.2 Requirements:

In order for the migration techniques to be effective and efficient, the migration solution must satisfy the following requirements:

1. Applications developed for and deployed on Grid systems are typically long running. Migration scheme adopted by the grid middleware should be strong migration in order to avoid to the lost of the work done till the time migration is initiated

2. Migration solution should exhibit transparency property. Application programmers must not be burdened by taking care of inserting the code for capturing and restoring the execution state.

3. Migrations scheme adopted must introduce least possible performance overhead.

4. Migration should be portable. No modification to the underlying operating system or other system software should be required

5. System should be able to initiate asynchronous migration. This is a very important requirement to enable load balancing

6. Migration solution should support multi-threaded application migration. Grid applications are mostly high performance applications and thus are multi-threaded.

## 4 DGET Middleware

Data Grid Environment & Tools (DGET)[15, 16, 17, 18] is peer-to-peer grid system being developed in UCD, Ireland. DGET exploits approaches from grid systems and p2p computing. Following are the major features of DGET middleware:

1. Transport protocol independent communication system

2. Decentralized P2P style resource discovery system

3. Uniform management interface

4. Resource accounting & control

5. Decentralized authentication mechanism
6. Policy based, dynamic and fine-grained access control
7. Simplicity & ease of use

## 4.1 DGET Concepts

**Entity:** Entity is the network enabled discrete unit of abstraction that provides some functionality to its users. Entity can take many forms e.g. a remote activity, a remote object, a server that processes user requests etc. Concept of an entity is akin to a process. Entity is a mobile element that can move around on different nuclei. Entity is composed of two parts, a system provided Shell and user provided Ghost. Definitions of these are given below.

**Shell** Shell is the system provided control part of the entity. Shell exposes a management interface through which entities can be manipulated. Shell is attached to the programmer provided Ghost when entity is instantiated.

**Ghost** Ghost represents the programmer provided part of an entity. Ghost implements the actual logic of the functionality.

**Nucleus** Nucleus is the kernel of the system. Provides basic services like lifecycle management, communication, security etc. to entities

## 4.2 Entity Execution Model

Entity instantiation is initiated by `EntityManager`. As described previously, entity is composed of two parts: `Shell` and `Ghost`. Therefore, entity instantiation procedure consists of two steps: instantiation of `Shell` and second, instantiation of the `Ghost`. In the first step, `EntityManager` instantiates the Shell passing the system parameters like class loaders to be used for loading Ghost classes and the ThreadGroup information. List of methods supported by Shell are given below.

```
public class Shell implements Runnable {
    public void Shell(
            DGETClassLoader cl, ThreadGroup tg){...}
    protected void start(String GhostMainClass){...}
    public void stop(){...}
    public void suspend(){...}
    public void resume(){...}
    public void export(java.net.URL destination){...}
    protected void import(ExecutionImage img){...}
}
```

The second part of entity instantiation is carried out by the `Shell`. `EntityManager` calls the `start` method on the shell passing the name of the main class of the Ghost. Main class of the programmer provided user logic must extend the `Ghost` class provided by DGET. `Shell` instantiates the `Ghost` and calls the `setEntityContext()` passing in the `EntityConext` and an instance of itself. `EntityContext` class represents the context in which the `Ghost` is running. Ghosts can get information about the host and resource consumption information through this. All the entity lifecycle related operations are invoked on the Shell which takes care of the operation invoked. Therefore, `Shell` passes an instance of itself so `Ghost` can invoke the lifecycle related events on itself through the `Shell`. For example, if a Ghost wants to migrate, it can invoke `export` method on its `Shell`. During the migration process, when execution state of an entity is to be restored, EntityManager calls `import` method instead of the `start` method.

## 5 Migration Support in DGET

This section describes the migration support present in the DGET middleware. Migration solution adopted in DGET takes into account all the requirements laid out in the previous section.

## 5.1 Implementation Methodology

Before going into details of the migration solution of DGET, we would like to give a brief overview of the implementation techniques used in the migration solution. Deciding factor in choosing these methodologies were the requirements of portability of the solution and minimal space & time performance overhead. Following two paragraphs explain the implementation techniques used:

**Bytecode Instrumentation** In order to support transparent migration in DGET, entity classes are instrumented and code blocks are inserted. These inserted code blocks perform different functions like program counter restoration, execution checkpoints (described shortly).This bytecode instrumentation is performed at class load time by a custom classloader. Bytecode instrumentation is performed by the classloader using the Byte Code Engineering Library(BCEL)[19]. BCEL provides an easy to use API for static analysis and dynamic creation or transformation of Java class files. It enables developers to implement the desired features on a high level of abstraction without handling all the internal details of the Java class file format.

**Java Platform Debugger Architecture (JPDA)** Standard JVMs don't expose any features to capture and restore the execution state of running threads. Existing migration solutions either rely on modified JVMs or instrumenting the class files with state capturing and restoration code. These approaches work well but result in the loss of portability or incur huge performance overhead due to the injected code for state capture and restoration.

We chose the Java Platform Debugger Architecture (JPDA)[20] to perform these tasks. JPDA is part of JVM specification and is implemented by every standard JVM implementation. JPDA provides access to runtime information of JVM including the thread stacks. JPDA is implemented purely in Java so our migration solution doesn't lose portability and state capture & restoration is performed with minimal performance overhead.

One drawback of using JPDA for state capturing and restoration prior to Java 1.4 was that when JVM was run in debug mode, JIT compilation was disabled and the programs were executed in interpreter mode. Beginning with Java 1.4, JVM runs at full speed in debug mode[21] Execution reverts to the interpreter mode during some debug operations such as single-stepping and when method entry/exit or watchpoints are set. We have ran some tests to execute programs in debug mode and normal mode to see the difference. This is a simple matrix multiplication program. Matrices of different sizes are generated in each iteration and the same method is called to multiply the matrices.

## 5.2 Migration Enabling Features

**Execution Checkpoints:** When the migration operation is initiated, entity threads can be at an arbitrary code location, for example in the middle of executing a source code level statement. At this point, there could be partial results on the operand stack. In order to perform migration at such point, values on the operand stack must be saved and restored during the entity restoration process. Unfortunately, JPDA doesn't expose any methods to access the operand stack. Initiating migration at such point would result in loss of data from the operand stack.

The solution to the problem mentioned in the previous paragraph could be to insert checkpoints in the code at locations where execution is not in the middle of an source code level statement. Migration requests should be delayed till the execution reaches any such checkpoint. When the thread reaches such execution checkpoint, if the execution status of the entity isn't suspended, thread can continue executing the entity code, otherwise, thread will be blocked. Execution checks are performed by making calls to `Execution` class. Execution status is maintained by an execution flag. `Check()` method call on the `Execution` class returns if the execution status of the entity is RUNNING, it blocks otherwise.

The questions that arises is how frequently and at what points these execution checks should be made. If execution checks are made too often, it would waste CPU power and degrade execution performance. On the other hand, infrequent execution checks would unnecessarily delay the initiation of migration process. Especially in the case of asynchronous migration initiated by the load balancer, migration process should begin as soon as possible. We decided to put these execution checks at the following locations:

- At the beginning of each method of all the entity classes. Operand stack is empty when the stack frame is created so initiating migration at this point will not result in any data loss from the operand stack.

- Within the loop body. At every iteration execution check will be performed before any code within the loop is executed. Putting execution checks in the loop body ensures that long loops wouldn't delay the initiation of migration operation. In the case of nested loops, execution checks are inserted within the body of innermost loop.

**Mobile Monitors:** One of the requirements laid out in section 2 was that the migration solution should be able to support multi-threaded migration. Multi-threaded migration adds a few complexities to the problem. In order to migrate multi-threaded application, locks acquired by threads must be preserved. Upon restoring the entity execution, lock state should also be restored and only thread holding the lock before migration should be allowed to run within the synchronized code. If lock state is not preserved, other thread might acquire the lock and start executing while the data protected by the locks could be in an inconsistent state. Java provides multi-threading support in the form of `Synchronized` methods and code blocks. A monitor is associated with each java object by JVM and before entering a `Synchronized` method or code block thread has to acquire the monitor associated with the object. Monitors associated with java objects are maintained and hidden inside the JVM. These monitors are not `Serilizable` and thus are not transported with the serialized objects.

In order to overcome this problem, we introduced Mobile Monitors in DGET. These mobile monitors are `Serilizable` and preserve the lock state upon migration. Upon serialization of these mobile monitors, the lock state is transferred as well. During the bytecode instrumentation process, Class constructors are instrumented and code is inserted to associate a mobile monitor with it. Standard Java Monitors are acquired and released with `monitorenter` and `monitorexit` bytecode instructions at the beginning and end of `Synchronized` code blocks. During the instrumentation, these instructions are replaced with method calls on the associated mobile monitor. Similarly, `wait()`, `notify()` and `notifyAll()` method calls are also replaced with calls on associated mobile monitor. As a result of this instrumentation, method call for lock acquisition and release are made on the mobile monitor associated with the object rather than the built in Java monitor.

## 5.3 Migration Process

**Entity Suspension:** Migration process is initiated when `export()` method is invoked on the `Shell`. Before the executions state can be captured, all the running threads of the entity must be suspended. Sun Microsystems has deprecated[22] `Thread.suspend()` and programmers are instructed to use alternative measures. Execution checkpoints discussed in the previous section are used to halt the execution of the entity. The `export` method calls `suspend()` method on the associated `Execution` class. As a result, execution of all the threads is blocked on the next execution checkpoint. At this point, all the entity threads would be in on of the following states:

1. Execution wait set: Thread reached execution checkpoint and blocked as a result to `check()` method call

2. Monitor entry set: Thread tried to acquire a lock and was blocked as lock was held by another thread

3. Monitor wait set: Thread called the wait() method inside a synchronized method or block and is blocked waiting for the `notify()` or `notifyAll()`

4. Thread called `sleep()` method and is sleeping

**State Capture:** After the execution of all the entity threads has been suspended, execution state capture can be started. JPDA discussed previously is used to capture the execution state of all the entity threads. `StackFrame` class from JPDA represents a method call on the thread stack. `StackFrame` class gives access to the values of local variables and the program counter. Each local variable is represented as `LocalVariable` class. Calling the `visibleVariable()` method on `StackFrame` class return a list of all the variables accessible till the point of execution in the method code. `location()` method return the `Location` class that represents the location in the stack frame. `codeIndex()` method can be used to extract the code index relative to the start instruction of the method. Using these classes and methods, execution state of all the methods on thread stack can be accessed and saved. Execution state of all the entity threads along with the mobile monitors and Execution class is saved in `Serializable` format and transported to the destination for reincarnation of the entity.

**State Restoration:** On the destination, entity state is restored by calling the `import()` method of the shell by the `EntityManager`. Saved image of entity's execution context is passed as parameter to the `import()` method. Entity threads must be launched in special order to avoid any race conditions. Threads blocked as a result of `wait()` method call must be launched first and then all the threads blocked on the `Execution` class. Threads in other states are started after them.

To reestablish execution state of a thread, its method stack must be rebuilt. To do this, all the methods are called in the order they were on the stack before execution was suspended and migration was initiated. JPDA allows to set event handlers which are called when method entry/exit event occurs. When a method entry event occurs, such event handler restores the values of local variable of the method from the saved execution image. After restoring local variables execution jumps to the code position which is method invocation for the next method on the stack. This execution jump is explained in the coming paragraph. Doing so would ensure the instructions already executed are skipped and restoration of next method on the stack frame begins and proceeds in the same manner. After restoring all the threads to the state they were before the migration was initiated, `resume()` method on the `Execution` class is called. This method sets the execution status flag to RUNNING and notifies all the threads blocked on this class. Execution will proceed normally afterwards.

As mentioned in the previous paragraph, after restoring local variables, execution jumps to the code position which is the method invocation of the next method on the stack. No mechanism is available in JPDA to set the value of the program counter register to this code position. This problem is solved by maintaining an artificial program counter (APC) which represents an index of method invocations in the method. This APC is incremented after every method invocation instruction. This APC is used in conjunction with a `tableswitch` bytecode instruction which branches the execution according the value of the APC. This `tableswitch` and APC increment instructions are added during the instrumentation process. `tableswitch` is added at the beginning of each method and defaults to the original starting code position of the method code.

For example the code block for a method given below is instrumented to add the APC support. There are 2 method invocations inside `method1()`. APC is maintained as a local variable and incremented after each method invocation.

**Source Code:**

```
public void method1(){
    ...
    method2();
    ...
    method3();
}
```

**Original Bytecode:**

```
3 aload_0
4 invokevirtual #31 <MyClass.method2>
8 aload_0
9 invokevirtual #36 <MyClass.method3>
18 return
```

**Modified Bytecode:**

```
0 iconst_0
1 istore_1
2 iload_1
3 tableswitch 0 to 2
       0:   8
       1:   15
       default:  4
7 aload_0
8 invokevirtual #31 <MyClass.method2>
9 iconst_1
10 istore_1 14 aload_0
15 invokevirtual #36 <MyClass.method3>
16 iconst_2
17 istore_1
26 return
```

Table 2. Execution Time

| App | Normal | Instrumented | Overhead |
|---|---|---|---|
| Simple | 328 ms | 359 ms | 31ms/9.45% |
| Complex | 604 ms | 657 ms | 53ms/8.77% |

Table 3. Comparative Evaluation

| Approach | Space Overhead | Execution Overhead |
|---|---|---|
| WASP | 400% | 20% |
| JavaGo | 280% | 183% |
| JavaGoX | 114% | 56% |
| Brakes | 107% | 27% |
| CIA | 10% | 814% |
| DGET | 15% | 9% |

## 6 Performance Evaluation

Instrumentation of entity classes add space and time overhead. Code blocks instrumented for execution checkpoints and program counter restoration increases the execution time of the entity. We ran some experiments to analyze this overhead. Two applications were ran to evaluate the solution. One is a very simple application to maintain a counter and display the value on the console. The other applications was a complex matrix multiplication. Experimental results are shown in the following two tables. Table 1 shows the space overhead caused due to the added code blocks in the entity classes. The second table shows the execution time of both the normal application and the instrumented application.

Table 1 shows results for space overhead caused by instrumentation of the entity classes. Simple application has 52 bytecode instruction before instrumentation and 60 after doing the instrumentation. The space overhead caused by instrumentation is 15%. In the second case, 10 more instructions were added and the overhead was 13% The second table shows the execution time of both application bother before and after performing instrumentation. The first application take 31ms extra to finish, thus causing 9% execution overhead. The execution overhead in the second application is the same 9%.

Table 3 shows numerical results of the existing approaches including DGET. The results for existing approaches are extracted from the respectively referenced paper. From the results, it can be seen that our approach has substantial performance advantage over the other approaches used for migration. In the case of CIA, reason for extra space overhead is due to the inserted execution checkpoints to enable asynchronous migration which is absent in CIA. CIA results were obtained on Java 1.3 which disables JIT in debug mode, therefore, it suffers from huge execution overhead. It would be hard to predict if the results will be same if Java 1.4 or later is used with CIA approach. In order to support asynchronous migration, some sort of approach like ours must be used which will result in the same extra overhead. Besides having quantitative advantage over other approaches, our mechanism support asynchronous migration as well as multi-threaded migration.

## 7 Conclusion and Future Work

In this paper, we presented mechanism to support transparent strong migration in grid environments. This mechanism is adopted and implemented in DGET middleware. Our migration solution is portable and incurs least possible space and time overhead. Migration can be self-initiated or can be invoked by some external entity like load balancer. Migration support for multi-threaded applications is also present which preserves the locks state of the entity threads. There are still some issues which remains unsolved, e.g. threads which are sleeping while migration is performed should not sleep more than the time initially set. The other topic of future work includes making the migration more fine-grained. Currently, migration can only be invoked while the thread is at the beginning of some method or within the loop. We will investigate ways to save and restore values on the operand stack so migration at any arbitrary code location can be performed.

Table 1. Space overhead (No. of Bytecode Instructions)

| App | Normal | Instrumented | Overhead |
|---|---|---|---|
| Simple | 52 | 60 | 15% |
| Complex | 151 | 171 | 13% |